\documentclass[useAMS,usenatbib]{mn2e}
\usepackage{psfig}
\usepackage{textcomp}
\usepackage[colorlinks=true,citecolor=blue]{hyperref}
\usepackage{graphicx,color}
\usepackage{dcolumn}
\usepackage{times}
\newcommand{\mnras}{MNRAS}

\title[INOV of NLSy1 galaxies]{Intranight optical variability of $\gamma$-ray 
loud Narrow Line Seyfert 1 galaxies}
\author[Vaidehi S. Paliya et al.] {Vaidehi S. Paliya $^{1,2}$
\thanks{E-mail: vaidehi@iiap.res.in},
C. S. Stalin$^{1}$, Brijesh Kumar$^{3}$, Brajesh Kumar$^{3}$, V. K. Bhatt$^{3}$, 
\newauthor
S. B. Pandey$^{3}$, R. K. S. Yadav$^{3}$ \\ 
$^{1}$Indian Institute of Astrophysics, Block II, Koramangala, Bangalore-560034, India\\
$^{2}$School of Inter-Disciplinary and Trans-Disciplinary Studies, IGNOU, New Delhi-110068, India \\
$^{3}$Aryabhatta Research Institute of Observational Sciences, Nainital-263129, India\\
}

\begin{document}

\date{Accepted 2012 October 13. Received 2012 September 16; in original form 2012 July 18}

\pagerange{\pageref{firstpage}--\pageref{lastpage}} \pubyear{2008}

\maketitle

\label{firstpage}

\begin{abstract}
The Large Area Telescope (LAT) onboard the {\it Fermi} Gamma Ray Space Telescope 
has detected $\gamma$-ray emission in about half-a-dozen Narrow Line Seyfert 1
(NLSy1) galaxies. This indicates the presence of relativistic jets in these 
sources similar to blazars and radio galaxies. In an attempt to have an idea
of the intranight optical variability (INOV) characteristics of these 
$\gamma$-ray loud NLSy1 galaxies, we have carried out optical flux monitoring
observations of three NLSy1 galaxies detected by {\it Fermi}/LAT: 1H 0323+342, 
PMN J0948+0022 and PKS 1502+036. These optical monitoring observations in 
R$_C$ band carried out during January $-$ May 2012, showed the presence of 
rapid optical flux variations in these sources. The intranight differential 
light curves
of these sources have revealed flux variations on time scales of hours
with amplitudes of variability $>$ 3 percent for most of the time. However,
for one source, PMN J0948+0022 we observed amplitude of variability 
as large as 52 percent. On using the {\it F}-statistics to classify the 
variability nature of these sources, we obtained a  
duty cycle (DC) of INOV of $\sim$85 percent. Alternatively, the more commonly 
used 
{\it C}-statistics gave a DC of INOV of $\sim$57 percent. Such high DC of INOV are
characteristics of the BL Lac class of AGN. The results
of our monitoring observations thus indicate that
there is similarity in the INOV nature of
$\gamma$-ray loud NLSy1 galaxies and BL Lac 
objects, arguing strongly for the presence of relativistic jets aligned 
closely to the observers line of sight in $\gamma$-ray loud NLSy1s. Moreover, 
our dense monitoring observations on some
of the nights have led to the clear detection of some mini-flares
superimposed on the flux variations during the night over 
timescales as short as 12 minutes. The detection of short timescale
flux variability in the sources studied here is clearly due to stronger 
time compression leading to the jets in these sources having large doppler 
factors, similar to that
of the inner jets of TeV blazars.

\end{abstract}

\section{Introduction}{\label{sec1}}
Flux variations across the electromagnetic spectrum is one of the defining
characteristics of Active Galactic Nuclei (AGN). They occur on a wide
range of timescales ranging from hours to days to months, making this
particular property of AGN an efficient tool to understand the physics
of these objects (\citealt{1995PASP..107..803U}; \citealt{1995ARA&A..33..163W}).
In AGN, where the flux is dominated by relativistic jets of non-thermal
emission pointed towards the direction of the observer, the observed
intensity variations will be rapid and of large amplitude (\citealt{1984RvMP...56..255B}). Such rapid and large amplitude variations, generally
explained by invoking relativistic jets (\citealt{1985ApJ...298..114M}; \citealt{1992ApJ...396..469H}; \citealt{1992vob..conf...85M}), have
been observed in the blazar class of AGN (\citealt{1990AJ....100..347C};
\citealt{1989Natur.337..627M}) mostly on hour like timescales and
recently on sub-hour timescales as well (\citealt{2010ApJ...719L.153R}; \citealt{2011ApJS..192...12I}).

\begin{table*}
\caption{List of the $\gamma$-NLSy1 galaxies  monitored in this work. Column 
informations are as follows: (1) IAU name; (2) other name; (3) right ascension 
in the epoch 2000; (4) declination in the epoch 2000; (5) redshift; 
(6) absolute B mangitude; (7) apparent V magnitude; (8) observed optical polarization; (9) radio 
spectral index and (10) radio loudness parameter 
R = f$_{1.4 GHz}$/f$_{440 nm}$(\citealt{2011nlsg.confE..24F})}.
\begin{tabular}{llcccccrrr}
\hline

IAU Name & Other Name  & RA (2000)$^a$ & Dec (2000)$^a$ & $z$$^a$ & M$_B$$^a$ & V$^a$ & P$_{opt}$ & $\alpha_R$$^d$ & R$^e$  \\
    &     &      &     &     & (mag) & (mag) & (\%) &     &  \\
(1) & (2) & (3)  & (4) & (5) &  (6)  &  (7)  &  (8) & (9) & (10) \\ 
\hline															     			
J0324+3410 & 1H 0323+342   & 03:24:41.2 & +34:10:45 & 0.063 & $-$22.2 & 15.72 & $<$ 1$^{b}$ & $-$0.35 & 318  \\
J0948+0022 & PMN J0948+0022 & 09:48:57.3 & +00:22:24 & 0.584 & $-$23.8 & 18.64 & 18.8$^{c}$  & 0.81    & 846 \\
J1505+0326 & PKS 1502+036  & 15:05:06.5 & +03:26:31 & 0.409 & $-$22.8 & 18.64 & ---         & 0.41    & 3364 \\ \hline
\end{tabular}

\hspace*{-11.0cm}$^a$ \citet{2010A&A...518A..10V}\\
\hspace*{-12.2cm}$^b$ \citet{2011nlsg.confE..49E}\\
\hspace*{-12.3cm}$^c$ \citet{2011PASJ...63..639I} \\
\hspace*{-0.9cm}$^d$ radio spectral index calculated using the 6 cm and 20 cm flux densities given in \citet{2010A&A...518A..10V} ($S_\nu \propto \nu^{\alpha}$) \\
\hspace*{-12.6cm}\noindent $^e$ \citet{2011nlsg.confE..24F}
\label{tab1}
\end{table*}
Narrow Line Seyfert 1 (NLSy1) galaxies are an interesting class of  
AGN similar to Seyfert galaxies that came to be known about
25 years ago by \citet{1985ApJ...297..166O}. Their optical spectra contain
 narrower than usual permitted lines from the Broad Line Region (BLR), having FWHM(H$_{\beta}$)
$<$ 2000 km sec$^{-1}$. Normally, they have [O~{\sc iii}]/H$_{\beta} <$ 3, however
exceptions are possible if they have strong [Fe~{\sc vii}] and [Fe~{\sc x}] 
lines(see
\citealt{2011nlsg.confE...2P} and references therein). Both BLR and
Narrow Line Region (NLR) are present in NLSy1s, but the permitted lines from BLR are narrower
than that of Seyfert 1 galaxies (\citealt{2000ApJ...538..581R}). 
NLSy1 galaxies were found to show rapid flux variability in the optical
(\citealt{2004ApJ...609...69K}; \citealt{2000NewAR..44..539M}).
They
also show the radio-loud/radio-quiet dichotomy (\citealt{2000ApJ...543L.111L}), however,
the radio-loud fraction of NLSy1 is about 7$\%$ (\citealt{2003ApJ...584..147Z};
\citealt{2006AJ....132..531K}), smaller than the fraction of 15$\%$ known in the
population of quasars (\citealt{1995PASP..107..803U}). These radio-loud NLSy1 
galaxies have flat radio
spectrum, high brightness temperatures, suggesting the presence of relativistic
jets in them (\citealt{2003ApJ...584..147Z}; \citealt{2006PASJ...58..829D}).

Since the launch of the {\it Fermi} Gamma-ray Space Telescope in 2008, high energy
(E $>$ 100 MeV) gamma rays were detected in a few radio-loud Narrow-Line Seyfert 1
galaxies. We therefore refer to these sources as $\gamma$-ray loud NLSy1 ($\gamma$-NLSy1) 
galaxies. These sources are found to have high energy 
variable $\gamma$-ray radiation as detected by {\it Fermi}/LAT. 
They are also found to have compact radio structure with a  core-jet 
morphology (\citealt{2006PASJ...58..829D}; \citealt{2011A&A...528L..11G}; 
\citealt{2007ApJ...658L..13Z};\citealt{2012arXiv1205.0402O}), superluminal 
motion and high brightness temperatures(\citealt{2012arXiv1205.0402O}; \citealt{2006PASJ...58..829D}).  
All these characteristics give a 
distinctive proof of the presence of relativistic jets in them 
(\citealt{2009ApJ...699..976A}; \citealt{2009ApJ...707..727A}; 
\citealt{2009ApJ...707L.142A}; \citealt{2010ASPC..427..243F}). 
Another independent proof of the existence of relativistic jets
oriented at small angles to the observer in these $\gamma$-NLSy1 
sources is the detection of intranight optical variability (INOV) on 
hour to sub-hour (minute) timescales.  
\citet{2010ApJ...715L.113L} reported the first detection 
of INOV in
the $\gamma$-NLSy1 galaxy PMN J0948+0022, wherein the authors 
detected INOV with amplitudes as large as 0.5 mag on 
timescale of several hours. Recently \citet{2011nlsg.confE..59M} also
found {\bf this} source to show INOV. In this work, we aim to understand the 
INOV characteristics of this new class of $\gamma$-NLSy1 galaxies, 
in particular to detect rapid INOV on these
sources on sub-hour (minute) time-scales and also to see
if their INOV characteristics compare with that of the 
blazar class of AGN known to have relativistic jets.  

We detail in Section \ref{sec2}, the sample of $\gamma$-NLSy1s selected for the 
intranight optical monitoring. Observations are described in 
Section \ref{sec3}. 
Section \ref{sec4} is devoted to the results of this work  
followed by discussion and conclusion in the final Section \ref{sec5}.


\section{Sample}{\label{sec2}}
One of the discoveries by {\it Fermi}/LAT was the detection of $\gamma$-rays
from the NLSy1 galaxy PMN J0948+0022 (\citealt{2009ApJ...699..976A}).
An analysis of the publicly available LAT data during the period
August 2008 to February 2011, has led to the detection of $\gamma$-rays 
in a total of seven NLSy1 galaxies including the source PMN J0948+0022
(\citealt{2011nlsg.confE..24F}). From this list of seven sources, we have selected for
this work only sources having significant detections in {\it Fermi}/LAT with the
test statistic (TS) larger than 100 and integrated $\gamma$-ray
flux in the 0.1$-$100 GeV range greater than 5 $\times$ 10$^{-8}$ 
photons cm$^{-2}$ s$^{-1}$. A TS value of 10 roughly corresponds
to 3$\sigma$ (\citealt{1996ApJ...461..396M}). This has led to the final 
selection
of three sources for intranight optical monitoring, namely
1H 0323+342, PMN J0948+0022 and PKS 1502+036.
The general properties of these
sources are given in Table \ref{tab1}.

\begin{figure*}
\vspace*{-6.0cm}
\hspace*{-1.0cm}\psfig{file=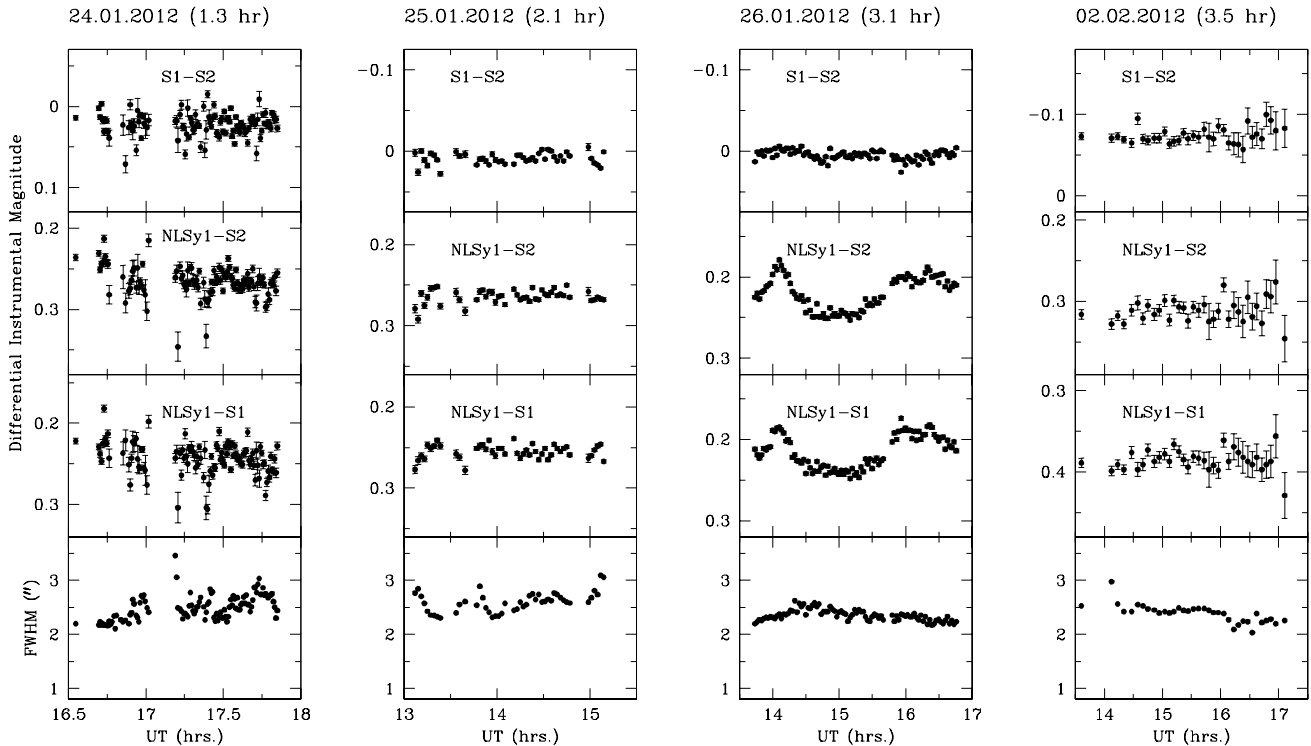}
\vspace*{-13.0cm}
\caption{Intranight DLCs of the source 1H 0323+342. The date
of observation and the duration of monitoring (within brackets)
are given on the top of each panel. On the bottom panel of each
night is given the variations of the FWHM of the stellar images
during the monitoring period in the night.}
\label{fig1}
\end{figure*}

\section{Observations and Reductions}{\label{sec3}}
Our observations were carried out on the newly commissioned 130 cm telescope 
(\citealt{2010ASInC...1..203S}) located at Devasthal and operated by the 
Aryabhatta Research Institute for Observational Sciences (acronym ARIES), 
Nainital. 
This telescope is a modified Ritchey Chretein system with a f/4 beam 
(\citealt{2010ASInC...1..203S}). We have used two detectors for our 
observations.  One is a 2k $\times$ 2k conventional CCD having a gain of 
1.39 e$^{-}$/ADU and read out noise of 6.14 e$^{-}$. Each pixel in this CCD 
has a dimension of 13.5 $\mu$m. This corresponds to 0.54 arcsec/pix on the sky 
thereby covering a field of 12 x 12 arcmin$^{2}$. The second detector used in 
our observation is the 512 $\times$ 512 Electron Multiplying Charged Coupled 
Device (EMCCD). 
It has very low readout noise (0.02 e$^{-}$) and a variable gain which can 
be selected by the observer. The preliminary science results from
observations of these CCDs are given by \citet{2012aj}. For observations reported 
here, we used a gain of 
225 e$^{-}$/ADU. It is well known from optical monitoring observations of 
blazars that the probability of finding INOV can be enhanced by continuous 
monitoring of a source (\citealt{1990BAAS...22R1337C}; 
\citealt{1995PhDT.........2N}) thus, in this work we tried to monitor each 
source continuously for a minimum of 4 hours during a night. However, due to 
weather constraints for some of the nights we were able to monitor sources for 
durations as low as 1 hour. All of the observations were done in 
Cousins R ($R_C$, hereafter)
filter as 
the CCD response is maximum in this band. The exposure time was typically 
between 30 seconds to 15 minutes depending on the brightness of the source, 
the phase of the moon and the sky transparency on that night. The target 
$\gamma$-NLSy1 galaxies were also suitably placed in the field of view (FOV), 
so as to have atleast three good comparison stars in the observed FOV.

Preliminary processing of the images, such as bias subtraction and flat 
fielding were done through standard procedures in IRAF\footnote [1]{IRAF is 
distributed by the National Optical Astronomy Observatories, which is operated 
by the Association of Universities for Research in Astronomy, Inc. under 
co-operative agreement with the National Science Foundation}. Aperture 
photometry on both the $\gamma$-NLSy1 and the comparison stars present on the cleaned 
image frames were carried out using the {\tt phot} task available within the
APPHOT package in IRAF. The optimum aperture radius for the photometry was 
determined using the procedure outlined in \citet{2004JApA...25....1S}. 
Firstly, star$-$star differential light curves were generated for a series of 
aperture radii starting from the median seeing FWHM of that night. The aperture
 that gave the minimum scatter for the different pairs of comparison stars was 
selected as the optimum aperture for the photometry of the target $\gamma$-NLSy1. 
Table \ref{tab2} consists of the positions and apparent magnitudes (taken from 
USNO\footnote[2]{http://www.nofs.navy.mil/data/fchpix/}) of the comparison 
stars used in the differential photometry. It should be noted that uncertainty 
in the magnitudes taken from this catalogue may be up to 0.25 mag. The log of 
observations are given in Table \ref{tab3}.

\section{Results}{\label{sec4}}
\subsection{Intranight optical variability}
\begin{figure*}
\vspace*{-6.0cm}
\hspace*{-1.0cm}\psfig{file=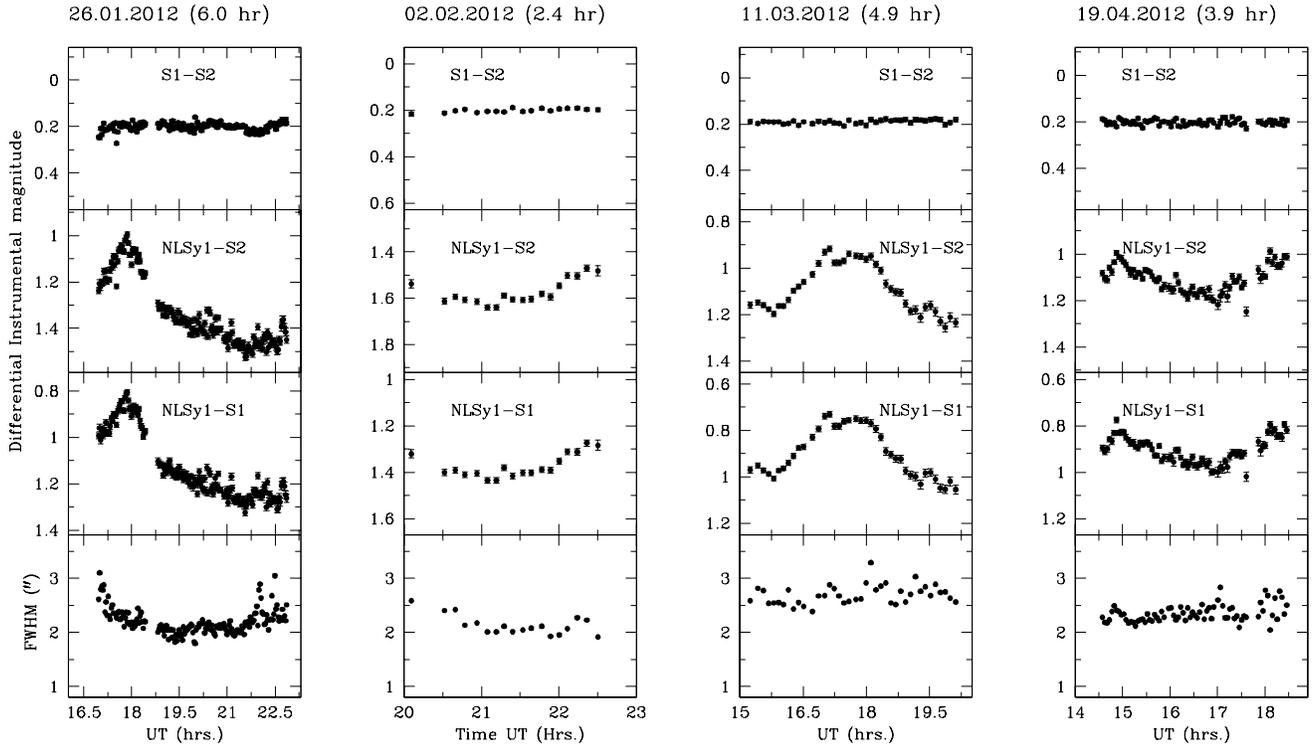}
\vspace*{-13.0cm}
\caption{Intranight DLCs for the $\gamma$-NLSy1 galaxy PMN J0948+0022. The 
variation of FWHM over the course of the night is given in the bottom panel.
Above the top panel, the date and duration of observations are given.}
\label{fig2}
\end{figure*}

\begin{figure*}
\vspace*{-6.0cm}
\hspace*{-1.0cm}\psfig{file=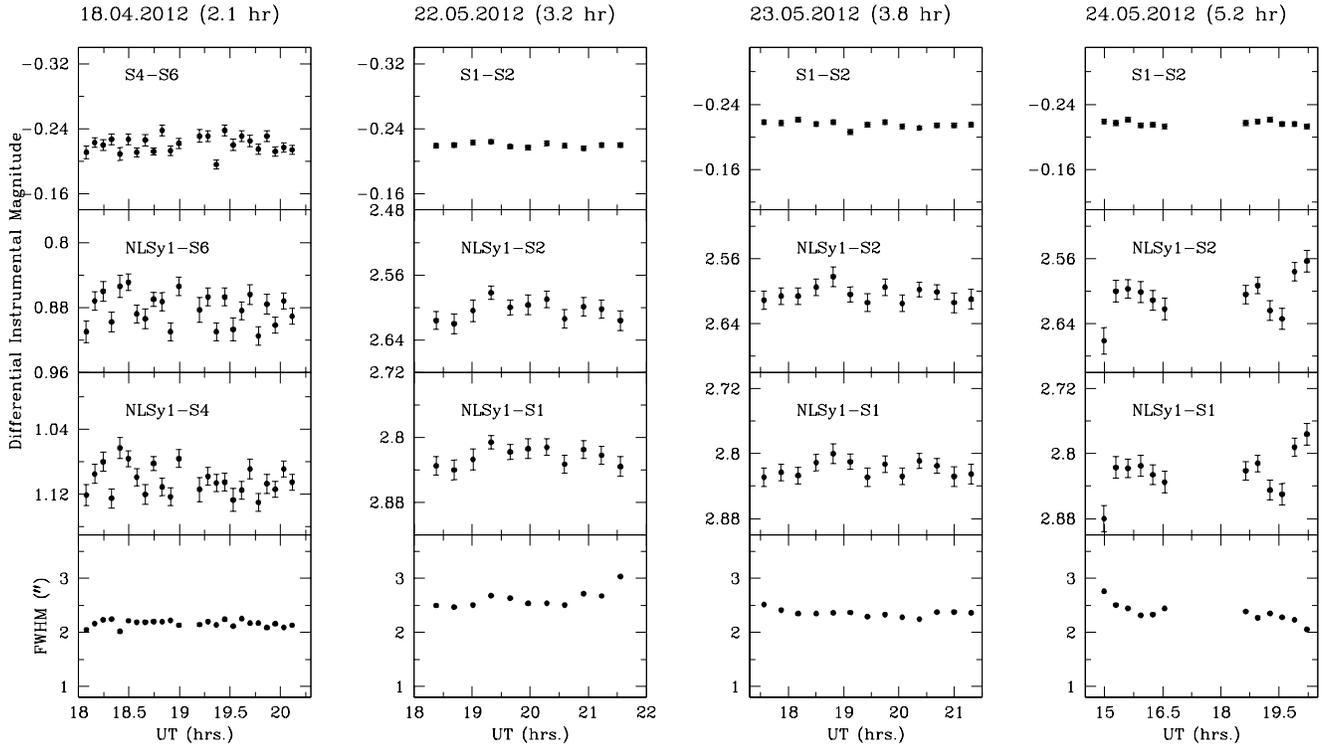}
\vspace*{-13.0cm}
\caption{Intranight DLCs for the $\gamma$-NLSy1 galaxy PKS 1502+036. The FWHM 
variation during the night is given in the bottom panel. The dates of observations
and the duration of monitoring are given on the top of each panel.}
\label{fig3}
\end{figure*}

From the derived instrumental magnitudes from photometry, differential light
curves (DLCs) were generated for the given $\gamma$-NLSy1 relative to steady comparison
stars. The optimum aperture used here is close to the median FWHM of the images
on any particular night most of the time. Also, as the central $\gamma$-NLSy1 dominates
its host galaxy, it should have negligible effects on the contribution 
of the host galaxy to the photometry (\citealt{2000AJ....119.1534C}). In
Figs.\ref{fig1}, \ref{fig2} and \ref{fig3} we present the DLCs for the objects 1H 0323+342, 
PMN J0948+0022 and PKS 1502+036 relative to two non-variable and 
steady comparison stars present in their observed frames.  
In order to judge the variability nature
of the sources, we have tried to use more than two comparison stars. However, in Figs \ref{fig1}, \ref{fig2} and \ref{fig3} we have shown the DLCs of $\gamma$-NLSy1
galaxies relative to only two comparison stars. These comparison stars were
 used later to characterise the variability of $\gamma$-NLSy1 galaxies. A  $\gamma$-NLSy1 is considered to be
variable only when it shows correlated variations both in amplitude and
time relative to all the selected comparison stars. Great care is taken 
on the selection of
comparison stars such that they are in close proximity to the source and
also have similar brightness to the source. However, it is not easy to get
such comparison stars, firstly  due to the small FOV covered by the 
EMCCD used in some nights of the observations reported here and
secondly due to the constraint of using the same standard stars
irrespective of which CCD was used for the observations. We note here that the uncertainty of the magnitudes of the 
comparison stars given in Table \ref{tab2} will have no effect on
the DLCs as they involve the differential instrumental
magnitudes between the $\gamma$-NLSy1 and the comparison stars. Also
the typical error of each point in the DLCs is around 0.01 mag. To 
access the variability nature of the sources, we have employed the 
following two criteria.  

\begin{table}
\caption{Positions and apparent magnitudes of the comparison stars from the USNO catalogue (\citealt{2003AJ....125..984M}).}
\begin{tabular}{lccccc}
\hline
Source & Star & RA(J2000) & Dec.(J2000) & R & B\\
 &  &  &  & (mag) & (mag)\\
\hline
1H 0323+342   & S1  & 03:24:44.26 & +34:09:31.84 & 15.17 & 14.82  \\
              & S2  & 03:24:51.24 & +34:12:26.53 & 15.06 & 14.91  \\
              & S3  & 03:24:35.01 & +34:09:36.73 & 15.81 & 15.43  \\
PMN J0948+0022 & S1  & 09:49:00.44 & +00:22:35.02 & 16.47 & 16.32  \\
              & S2  & 09:48:57.30 & +00:24:18.53 & 16.15 & 16.35  \\
              & S3  & 09:48:53.69 & +00:24:55.14 & 16.14 & 16.34  \\
 PKS 1502+036 & S1  & 15:05:11.30 & +03:22:25.57 & 15.24 & 16.20  \\
              & S2  & 15:05:15.90 & +03:19:10.91 & 15.44 & 16.25  \\
              & S3  & 15:05:26.99 & +03:24:37.25 & 15.71 & 16.95  \\
              & S4  & 15:05:15.37 & +03:25:40.90 & 17.12 & 18.85  \\
              & S5  & 15:05:15.65 & +03:25:27.62 & 16.55 & 17.92  \\
              & S6  & 15:05:14.53 & +03:24:56.34 & 16.74 & 17.08  \\
\hline
\end{tabular}
\label{tab2}
\end{table}

\subsubsection{C-Statistics}
To decide on the INOV nature of the sources on any given night of 
observations we have used the commonly used statistical criteria 
called the {\it C} parameter.  
Follwing (\citealt{1997AJ....114..565J}), for any given DLC, it 
is defined as 

\begin{center}
 \begin{equation}\label{eq1} 
C = \frac{\sigma_{T}}{\sigma},
\end{equation}
\end{center}

where $\sigma_{T}$ and $\sigma$ are the standard deviations of the source 
and the comparison star differential light curves. As we have used
three comparison stars, we have three star$-$star DLCs. Of these, 
we consider that DLC where the standard deviation of the light curve
is minimum, as this will involve the steadiest pair of comparison stars.
The $\sigma$ of this steadiest DLC is used in Eq. 1, to get two estimates
of the {\it C}-statistics for the source$-$star DLC, corresponding
to each of the comparison stars. A source is considered 
to be variable if {\it C} $\geq$ 2.576, which
corresponds to a 99$\%$ confidence level (\citealt{1997AJ....114..565J}).
Here, we get two values of {\it C}, corresponding to two DLCs of the source
relative to each of the two comparison stars. Using {\it C} statistics, 
we consider a source to be variable, when both the calculated 
{\it C} values exceed 2.576.

\subsubsection{F-statistics}
Recently, \citet{2010AJ....139.1269D} has argued that the {\it C}-statistics 
widely used in characterising AGN variability is not a proper statistics
and is wrongly established. An alternative to {\it C}-statistics according to 
\citet{2010AJ....139.1269D}, which can better access the variations in 
AGN light curves is the {\it F}-statistics. This statistics takes into 
account the ratio of 
two variances given as
\begin{equation}
F = \frac {\sigma^2_T}{\sigma^2}
\end{equation}
where ${\sigma^2_T}$ is the variance of the source$-$comparison star DLC and 
$\sigma^2$ is the variance of the comparison stars DLC. Thus, in computing
the {\it F}-statistics, of the three comparison star DLCs, similar to
the calculation of {\it C}-statistics,  we have selected the
comparison stars DLC that involves the steadiest pair of comparison stars.

Using Eq. 2, for each source, two values of the  {\it F} statistics were 
computed for the source$-$star DLCs corresponding to each of these two comparison
stars.  This calculated {\it F}-statistics is then compared with the critical 
F-value, F$^{\alpha}_{\nu}$, where $\alpha$ is the significance level 
and $\nu$ (= N$_{p}$ $-$ 1) is the degree of freedom for the DLC. We have 
used a significance level of $\alpha$ = 0.01, corresponding to a confidence 
levels of p $>$ 99 percent.  Thus, if both the computed {\it F}-values, 
corresponding to the DLCs of the source to each of the two comparison stars
are above the critical {\it F}-value corresponding to p $>$ 0.99, we 
consider the source to be variable. However, we point out that the
{\it C} statistics might be a more compelling measure of the presence
of variability, particularly when the comparison star light 
curves are clearly not steady.
 Also, the variations in the FWHM of the point sources in the
observed CCD frames, during the course of the night might give rise to
fictitious variations in the target NLSy1 galaxies. However, the variations 
of the NLSy1s detected in the observations reported here do not have any 
correlation with the FWHM variations and are thus genuine variations 
of the NLSy1s.

\subsubsection{Amplitude of Variability ($\psi$)}
The actual variation displayed by the $\gamma$-NLSy1 galaxies on any given night
is quantified using the INOV variability amplitude after correcting for the
error in the measurements. This amplitude, $\psi$ is determined using the
definition of \citet{1999A&AS..135..477R}

\begin{equation}\label{eq6} 
\psi = 100 \sqrt{D_{max} - D_{min})^{2} - 2\sigma^{2}}\%
\end{equation}
with \\
D$_{max}$ = maximum in $\gamma$-NLSy1 differential light curve,
\\ D$_{min}$ = minimum in $\gamma$-NLSy1 differential light curve,
\\ $\sigma^{2}$ = variance in the star$-$star DLC involving the steadiest pair
of comparison stars \\
The amplitude of variability calculated using Eq. \ref{eq6} for 
the nights when INOV was observed is given in Table \ref{tab4}.
The results of INOV are also given in Table \ref{tab4}. Also, we 
mention that, given the random variability nature of the sources with occasional short time scale and large 
amplitude flares, it is possible to detect large amplitude variability in 
these NLSy1s, if they are monitored for a longer duration of time.
\subsubsection{Duty Cycle}
Definition of duty cycle (DC) from \citet{1999A&AS..135..477R} was used to 
calculate it in our observations. 
It is very well known that objects may not show flux variations on all the 
nights they were monitored. Therefore, it is appropriate if DC is
evaluated by taking the ratio of the time over which the
object shows variations to the total observing time, instead 
of considering the fraction of variable objects.
Thus, it can be written as

\begin{equation}\label{eq7}
DC  = 100\frac{\sum_{i=1}^n N_i(1/\Delta t_i)}{\sum_{i=1}^n (1/\Delta t_i)} {\rm ~\%}
\end{equation}

where $\Delta t_{i}$ = $\Delta t_{i,obs}(1 + z)^{-1}$ is the duration of the 
monitoring session of a source on the {\itshape i}$^{th}$ night, corrected for 
its cosmological redshift $z$. 
N$_{i}$ was set equal to 1 if INOV was detected, otherwise N$_{i}$ = 0.
Using only the {\it C}-statistics to judge the presence of INOV, on any 
particular night, we find a DC of 57 percent for INOV of these 
$\gamma$-ray NLSy1 galaxies. However, 
using the {\it F}-statistics, we find
an increased DC of INOV of 85 percent.
\subsection{Long term Optical Variability (LTOV)}
Since the total time-span covered for observations range from days to months, 
this allowed us to search for the long term optical variability (LTOV) of the sources.
The LTOV results are summarized in the last column of Table \ref{tab4}. Here, 
the values indicate the difference of the mean $\gamma$-NLSy1 DLC with respect 
to the {\bf previous} epoch of observation.

\begin{table}
\caption[Log of observations]{Log of observations. Columns:- (1) source name; (2) date of observations; 
(3) duration of monitoring in hrs.; (4) number of data points in DLC; 
(5) exposure time in second; (6) CCD modes used; here "normal" refers to
the 2k $\times$ 2k pixels$^2$ CCD and "EM" refers to the 512 $\times$ pixels$^2$ EMCCD}
\begin{tabular}{lccccc}
\hline
Source & Date & Duration & N & Exp.time & CCD mode\\
       &      & (hrs.)   &      & (secs.)  & \\
 (1)   & (2)  &   (3)    &  (4) &  (5)     & (6) \\
\hline
1H 0323+342 & 24.01.12 & 1.3 & 105 & 30 & EM\\
 & 25.01.12 & 2.1 & 47 & 120 & EM\\
 & 26.01.12 & 3.1 & 89 & 120 & EM\\
 & 02.02.12 & 3.5 & 35 & 150 & Normal\\
PMN J0948+0022 & 26.01.12 & 6.0 & 165 & 120 & EM\\
 & 02.02.12 & 2.4 & 18 & 200 & Normal\\
 & 11.03.12 & 4.9 & 40 & 150 & Normal\\
 & 19.04.12 & 3.9 & 73 & 180 & EM\\
PKS 1502+036 & 18.04.12 & 2.1 & 24 & 300 & EM\\
 & 22.05.12 & 3.2 & 11 & 900 & Normal\\
 & 23.05.12 & 3.8 & 13 & 900 & Normal\\
 & 24.05.12 & 5.2 & 12 & 900 & Normal\\
\hline
\end{tabular}
\label{tab3}
\end{table}

\begin{table*}
\caption[Log of INOV and LTOV observations]{Log of INOV and LTOV observations. Columns:- (1) source name; (2) date of observation; 
(3) INOV amplitude in percent; (4)  and (5) {\it F}-values computed for 
the $\gamma$-NLSy1 galaxy DLCs relative to the steadiest pair of comparison 
stars on any night; (6) variability status according to {\it F}-statistics; 
(7) and (8) Values of {\it C} for the two $\gamma$-NLSy1 galaxy DLCs relative 
to the two comparison stars; (9) variability status as per {\it C}-statistics 
and (10) magnitude difference for LTOV relative to the first epoch of observation.} 
\begin{center}
\begin{tabular}{lcrrrcrrcr}
\hline 
Source & Date     & $\psi$ & {\it F1}  & {\it F2} & Status  & {\it C1}  & {\it C2}  &  Status & LTOV        \\
       & dd.mm.yy & ($\%$) &     &    &         &     &     &         & ($\Delta$m) \\
 (1)   & (2)      & (3)    & (4) & (5)& (6)     & (7) & (8) & (9)     &  (10)        \\
\hline
1H 0323+342 & 24.01.12   &  12.69  &  1.903  & 1.736  & V   & 1.380  & 1.318  &  NV  &            \\
            & 25.01.12   &  ----   &  1.414  & 1.357  & NV  & 1.189  & 1.165  &  NV  & $-$0.01   \\
            & 26.01.12   &  7.35   & 12.445  &12.406  & V   & 3.528  & 3.522  &  V   &    0.04   \\
            & 02.02.12   &  ----   &  1.766  & 2.153  & NV  & 1.329  & 1.467  &  NV  & $-$0.17   \\
\\
PMN J0948+0022 & 26.01.12 &  51.92  & 114.427 &107.992 & V   & 10.697 & 10.392 &  V  &             \\
              & 02.02.12 &  17.12  &  54.448 & 56.270 & V   &  7.411 &  7.501 &  V  &  $-$0.24   \\
              & 11.03.12 &  33.13  & 168.340 &161.495 & V   & 12.975 & 12.708 &  V  &     0.47   \\
              & 19.04.12 &  25.25  &  24.071 & 25.500 & V   &  4.906 &  5.050 &  V  &  $-$0.01   \\
\\
PKS 1502+036 & 18.04.12  &  6.49   &  3.263 &  3.913  & V   & 1.806  & 1.978  &  NV  &             \\
             & 22.05.12  &  3.58   & 21.744 & 23.753  & V   & 4.663  & 4.874  &  V  &             \\
             & 23.05.12  &  3.16   &  8.429 & 10.186  & V   & 2.903  & 3.192  &  V  &             \\
             & 24.05.12  & 10.09   & 93.043 & 86.512  & V   & 9.646  & 9.301  &  V  &             \\
 \hline
\end{tabular}
\label{tab4}
\end{center}
\end{table*}

\begin{figure*}
\includegraphics{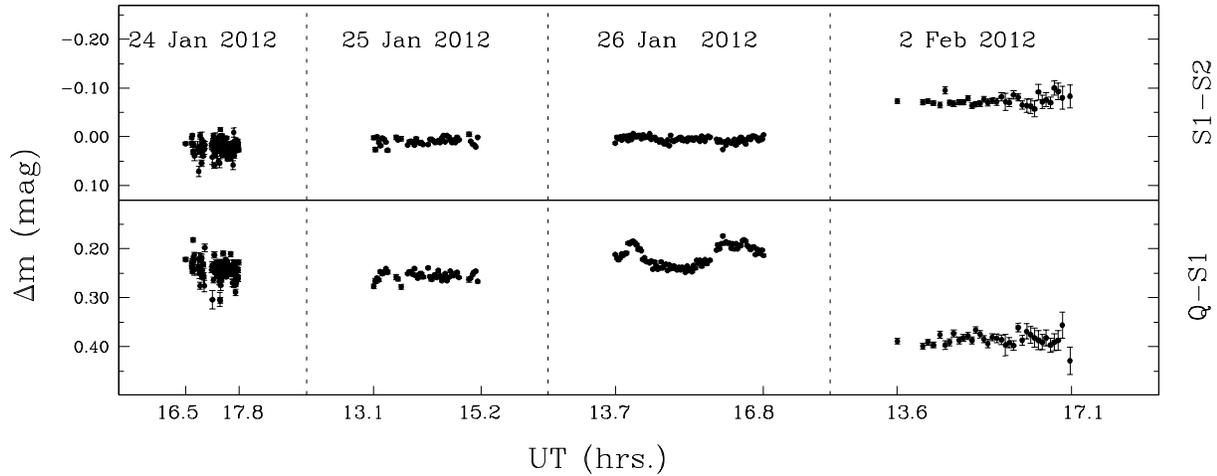}
\caption{Long term light curves of 1H 0323+342}
\label{fig4}
\end{figure*}

\section{Notes on the individual sources}{\label{sec5}}
\subsection{1H 0323+342}
This source was found to be a $\gamma$-ray emitter 
by {\it Fermi}/LAT(\citealt{2010ApJ...715..429A}).
\citet{2011nlsg.confE..49E} found optical polarization $<$ 1\% on Feb. 07 and
Feb. 10, 2011.  HST images
were well decomposed using a central 
point source component above a S\'{e}rsic profile (\citealt{2007ApJ...658L..13Z}). 
VLBI observations give clear evidence of a core-jet 
structure for this source ({\citealt{2007ApJ...658L..13Z}). Using the 6 cm and 20 cm flux densities given 
in the \citet{2010A&A...518A..10V} catalogue, the source is found to have an
flat radio spectrum with $\alpha_R$=$-$0.35 ($S_{\nu} \propto \nu^{\alpha}$).
This source has not been studied for INOV prior to this work.
Observations with good time resolution was obtained for a total of four
nights. To judge the variability of 1H 0323+342, we have
selected three comparison stars, namely S1, S2 and S3. However, as S3 was
clearly non-steady, it was not used for generating the DLCs of
1H 0323+342.  
On 24.01.2012, the comparison stars were not found to be 
stable. On this night, the observations have an average 
time resolution of 30 seconds. From {\it F}-statistic we find  
the $\gamma$-NLSy1 DLCs to be variable. However, from {\it C}-statistics, 
the $\gamma$-NLSy1 galaxy have a {\it C} $<$ 2.576,
making it to be non-variable on that night. 
One day later, on 25.01.2012, 
the source was again observed for a total duration of
2.1 hours with a typical sampling of 1 data point every 2 min. 
The source was non-variable on this night using
both the {\it C} and {\it F} statistics.
On 26.01.2012, about 3.5 hours of data
were gathered on the source with a temporal resolution of 
2 min, and the source was found to show 
clear variations on that night. It was also found to be non-variable
over the 5 hours of monitoring done on 02.02.2012. On this 
night we have on average one data point around every 5 min.
Thus, the source was found to show unambiguous evidence
of INOV on one of the 4 nights of monitoring. However,
on 24.01.2012, according to {\it F}-statistics, the source is
variable, while it is not so, when {\it C}-statistics was used.
Considering the LTOV of the source, the total 
time baseline covered for this source
is 8 days. Over the course of 8 days, the source was found to show 
variations. It faded by 0.01 mag in the first 24 hours, however
brightened by 0.04 mag in another 24 hours  and again became
fainter by 0.17 mag over 6 days between 26.01.2012 and 02.02.2012 (Figure \ref{fig4}).

\begin{figure*}
\includegraphics{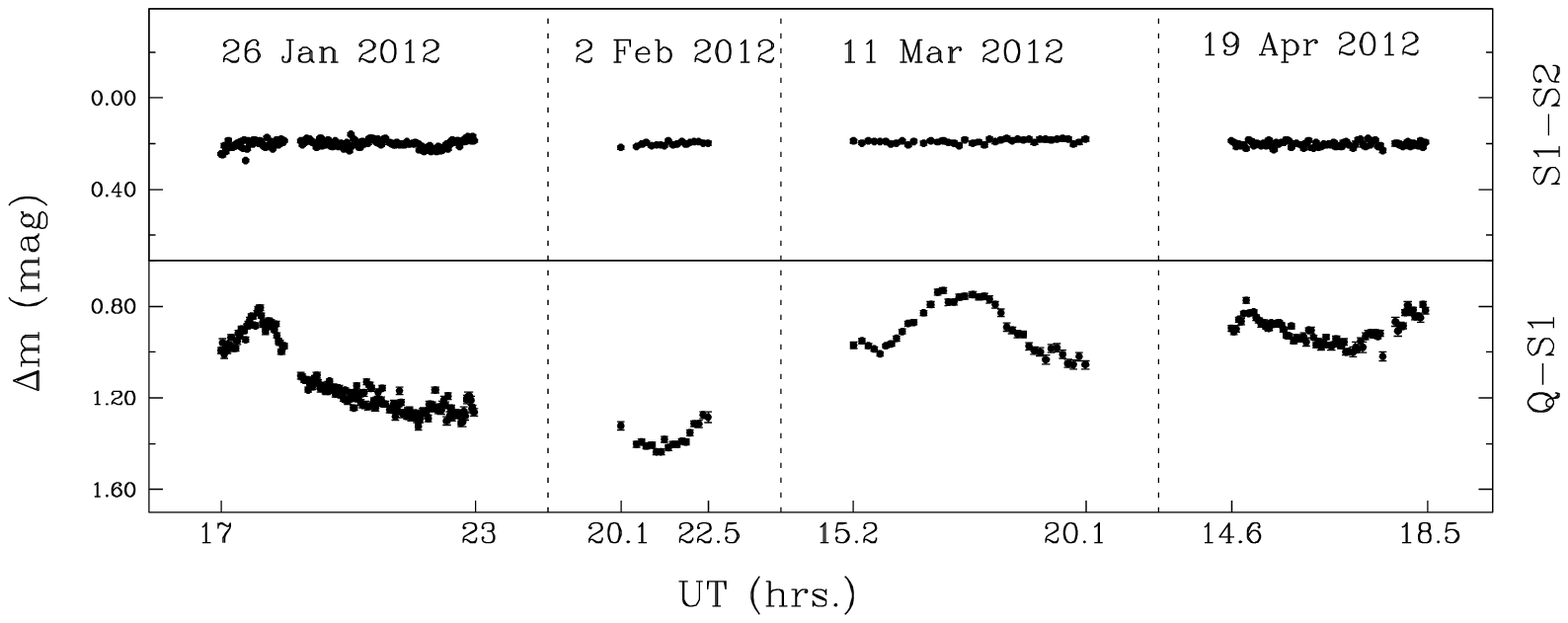}
\caption{Long term light curves of PMN J0948+0022}
\label{fig5}
\end{figure*}

\subsection{PMN J0948+0022}
This was the first NLSy1 galaxy detected in the $\gamma$-ray band during
the initial months of operation of {\it Fermi} (\citealt{2009ApJ...707..727A}).
It was found to have high optical polarization of 18.8\% when observed
during March$-$April 2009, by \citet{2011PASJ...63..639I}. However, it showed
low optical polarization of about 1.85\% when observed
again on Feb. 10, 2011 (\citealt{2011nlsg.confE..49E}). Such polarization variations
are not uncommon in  blazars.
It has an inverted radio spectrum with $\alpha_R$ = 0.81 evaluated
using the 6 cm and 20 cm flux densities given in 
\citet{2010A&A...518A..10V} catalogue.
VLBI observations revealed high brightness temperature and a compact
structure (\citealt{2006PASJ...58..829D}), with a possible core-jet
morphology (\citealt{2011A&A...528L..11G}). 
Previous INOV observations showed the source to show violent variations
with amplitudes of variability as large as 0.5 mag in timescale
of hours (\citealt{2010ApJ...715L.113L}). Recently, 
\citet{2011nlsg.confE..59M} has also detected INOV in PMN J0948+0022. 
 This source
has been monitored by us for four nights with durations ranging from two to 
six hours between Jan to Apr 2012. To characterise the variability
of PMN J0948+0022 during the nights it was monitored, we have selected
three comparison stars, S1, S2 and S3, all of which were found to 
be non-variable. However, for all variability analysis we 
have considered only the DLCs of PMN J0948+0022 relative to S1 and S2.
On 26.01.2012, it was monitored for a total duration
of 7 hours, with a good time resolution of about 2 minutes. 
Clear evidence of variability was found on this night with
amplitudes of variability as large as 52\%. A fast increase in brightness
to 0.1 mag and slow declining flare with peak at $\sim$17.9 UT was
found . The source then displayed a gradual decrease in brightness during 
the course of the night.  Superimposed on this brightness change of the
source, we found several mini-flares with timescales as short
as 12 min. One such mini-flare is towards
the end of the night. Between 22.6 and 22.9 UT, the source
increased in brightness by 0.12 mag, reaching maximum brightness 
at 22.76 UT and then gradually decreased in brightness.
These mini-flares are in fact real and cannot be attributed 
to seeing fluctuations during the course of the night, as we do not see 
any correlation between the occurrence of the mini-flares and fluctuations
in the FWHM variations. The FWHM after 22.5 UT was nearly steady, whereas
a brightness change of 0.12 mag was noticed in the $\gamma$-NLSy1.
On the observations done on  02.02.2012 for a duration of 2.4 hours with a temporal resolution of about 7 min,
the source showed a gradual brightness change of about 0.2 magnitude 
during the course of our observations. A large flare over a period
of 3 hours was found during the 4.9 hours of observations on 11.03.2012. On this night, the average time resolution of the observations 
is about 7 min.
Again on this night, superimposed on the large flare we noticed two mini-flares
one at 17  UT and the other at 19.5 UT. The mini-flare at 19.5 UT
showed a fast increase in brightness by 0.05 mag between 19.28 and 
19.52 UT and then gradually reached the original brightness level
at 19.87 UT. The total flare duration is $\sim$35 min. with a rise
time of $\sim$14 min. and a decay time of $\sim$21 min.
During 17 UT the FWHM has become poorer by 0.2$^{\prime\prime}$, 
whereas the $\gamma$-NLSy1 galaxy increased in brightness by 0.05 mag.
The increase in brightness of the NLSy1 at 19.5 UT is not 
associated with the FHWM becoming poorer by 0.2$^{\prime\prime}$, as we
might expect the source to become fainter due to FWHM degradation. 
Thus, the two mini-flares observed
on this night are also real and they are not due to any changes in the seeing
variations during those times.
INOV was also detected on the observations done on
19.04.2012. On this night, the light curve is densely sampled
wherein we have on average one data point every 3 min. Thus, the source has shown variations on all the four nights monitored by us.
The LTOV of this source can be noticed from the four epochs of monitoring over a duration of
4 months. Between the first two epochs, separated by six days, the source
has decreased in brightness by about 0.2 mag. It then brightened by $\sim$0.5 mag between
02.02.2012 and 11.03.2012, and again became fainter by about 0.01 mag when observed
on 19.04.2012 (Figure \ref{fig5}).

\subsection{PKS 1502+036}
This source was found to be  emitting in $\gamma$-ray 
band by {\it Fermi}/LAT (\citealt{2010ApJ...715..429A}) and
is the faintest $\gamma$-ray loud NLSy1 known in the northern 
hemisphere as of now.
It was found to have a core-jet structure from VLBA 
imaging (\citealt{2012arXiv1205.0402O}).
Its radio spectrum is inverted with $\alpha_R$ = 0.41.
PKS 1502+036 was monitored by us on four nights for INOV. 
We have used six comparison stars which are brighter than 
the source PKS 1502+036, to detect the presence 
of variability in it mainly due to the non-availability of 
suitable comparison stars of brightness similar to PKS 1502+036 in 
the observed field. For characterization of variability either using
the {\it C} statistics or {\it F} statistics, we have used
the stars S4 and S6 for the observations of 18.04.2012, whereas, for
the other three nights we have used the stars S1 and S2. The three
nights of observations carried out in May have an average time 
resolution of 19 min, whereas on 18.04.12, we have a better 
sampling of one data point every 5 min. On the observations
done on 18.04.2012, INOV could not be detected. 
Clear INOV was also detected when the source was monitored
on 22.05.2012 for a duration of 3.2 hours, though  the order of
fluctuations in magnitude was found to be very small. A gradual increase in
brightness of 0.03 mag over a period of 2 hours and then a decrease by 0.035
mag in the next one and a half hour was found. Source showed the largest variability 
on 24.05.2012,  with amplitude of variability as large as $\sim$10 percent. 
The observations made on this source covers a total time baseline of 
about a month. As the comparison stars used during the April and May
observations were different, the LTOV during this period could not be 
ascertained. However, from the observations done in May, the source
brightness remained the same both on 22 and 23 May, but faded by $~$0.045
mag when last observed on 24.05.2012. The large error bars 
in the DLCs of PKS 1502+036 is mainly due to its faintness. Though from
visual examination it is difficult to unambiguously identify the variations, 
using the conservative {\it C} statistics we  
classify the source to be  variable on three of the four nights 
it was monitored.

\section{Conclusions and Discussion}{\label{sec6}}
All the three sources of this present study have flat/inverted radio spectra,
and also show $\gamma$-ray flux variability (\citealt{2003ApJ...584..147Z},
\citealt{2009ApJ...699..976A}, \citealt{2009ApJ...707..727A}). The source 
PMN J0948+0022
showed a large polarization of 18.8\% in 2009 (\citealt{2011PASJ...63..639I}) 
however, was found to be in a low polarized state with P$_{opt}$ of 1.85\%
in 2011 (\citealt{2011nlsg.confE..49E}). Optical polarization of $<$ 1\% was also
observed for the source 1H 0323+342 (\citealt{2011nlsg.confE..49E}). High resolution 
radio observations of these $\gamma$-NLSy1 galaxies
point to the presence of 
core-jet structure, superluminal motion and high brightness temperatures
(\citealt{2006PASJ...58..829D}; \citealt{2011A&A...528L..11G}; 
\citealt{2007ApJ...658L..13Z}; \citealt{2012arXiv1205.0402O}).
All these observations clearly point to the presence of
relativistically beamed jets in these sources, closely aligned to the
observers line of sight.
It is known that $\gamma$-ray detected blazars
show more INOV, compared to non-$\gamma$-ray detected blazars pointing to
an association between INOV and relativistic jets, that are more aligned
to the observers line of sight (\citealt{2011MNRAS.416..101G}). 
Earlier studies have shown that 
large amplitude $\psi$ $>$ 3\% and high DC of around 70\% is exhibited
by the BL Lac class of AGN (\citealt{2004JApA...25....1S}; \citealt{2004MNRAS.348..176S}).

The spectral energy distribution (SED) of $\gamma$-NLSy1 galaxies is found 
to be similar
to that of blazars (\citealt{2012arXiv1207.3092D};\citealt{2009ApJ...707..727A}
;\citealt{2011MNRAS.413.1671F}). This non-thermal
continuum spectra consists of distinct low energy (due to synchrotron emission
mechanism) and high energy (due to inverse Compton emission mechanism)
components. The polarized optical flux seen in $\gamma$-NLSy1 galaxies
(\citealt{2011nlsg.confE..49E};\citealt{2011PASJ...63..639I})
is therefore a manifestation of relativistically beamed synchrotron emission
which also accounts for the low energy component of their SED,    
similar to the blazar class of AGN. From high resolution
VLBI observations and optical monitoring data of AGN, optical flare 
rise is  always associated with the emergence of new superluminal 
blobs of synchrotron plasma (knots)  in the 
relativistic jet (\citealt{2010ApJ...715..355L};
\citealt{2010MNRAS.401.1231A}). Correlations between flux and polarization 
variations
were observed in blazars such as Mrk 421 (\citealt{1998A&A...339...41T}) 
and AO 0235+164 (\citealt{2008ApJ...672...40H}). Similar to blazars, 
the flux variations in $\gamma$-NLSy1 galaxies
can be
explained by the shock-in-jet model (\citealt{1985ApJ...298..114M}). 
The turbulent jet
plasma when it passes through the shocks in the jet downstream could give
rise to increased multiband synchrotron emission  and 
polarization (\citealt{2012A&A...544A..37G}
and references therein). Recently, it has been found by
\citet{2012A&A...544A..37G}, that sources with strong optical polarization also show 
high INOV.

Though there are ample observational evidence for the presence of closely
aligned relativistic jets in these $\gamma$-NLSy1 galaxies, an
independent way to test their presence is to look for INOV in them.
The prime motivation for this work is therefore,
to understand the INOV characteristics of this new class of $\gamma$-ray loud
NLSy1 galaxies and also to see for similarities/differences with respect to the
$\gamma$-ray emitting blazar population of AGN. 
The observations presented here report the  
INOV characteristics of the sample of three $\gamma$-NLSy1 galaxies. 
The sample of sources in the present study consists of three out of the
7 known $\gamma$-ray loud NLSy1 galaxies, and therefore the INOV results 
found here
might be representative of the INOV characteristics of the new population
of $\gamma$-ray loud NLSy1 galaxies. 
Our 
high temporal sampling observation carried out on some of the nights using
the EMCCD, have enabled us to detect ultra rapid continuum flux variations
in the source PMN J0948+0022. Such rapid flux variations are possible as it is
known that the jet in $\gamma$-ray bright AGN have large bulk flow Lorentz
factor, thereby oriented at small angles to the line of sight leading
to stronger relativistic beaming (\citealt{2009A&A...507L..33P}).

From the observations of 3 sources, over 10 nights, using {\it C}-statistics
 we find a DC of 
variability of 57 percent. However, this increased to 85 percent 
when the {\it F}-statistics
discussed in \citet{2010AJ....139.1269D} was used. Also, the amplitudes of 
variability 
($\psi$) is found to be greater than 3\% for most of the time. Such high amplitude ($\psi >$ 3\%), 
high DC ($\sim$ 70 percent) INOV are
characteristics of the BL Lac class of AGN (\citealt{2004JApA...25....1S})
and thus we conclude that the INOV characteristics of $\gamma$-NLSy1 galaxies
are similar to blazars.
The present study therefore,
provides yet another independent argument for the presence of relativistic
jets in these $\gamma$-ray loud NLSy1 galaxies closely aligned
to the observer similar to the blazar class of AGN. 

Our present observations also indicate that $\gamma$-ray loud NLSy1 galaxies
do show LTOV on day to month like time scales, similar to that
shown by other classes of AGN (\citealt{2000ApJ...540..652W}; \citealt{2004JApA...25....1S} ).
However, due to the limited nature of our observations, with each sources
observed over different time baselines, definitive estimates of the 
LTOV of these sample of sources could not be made.
Though there are ample evidences for the presence of jets in these
sources, both from the INOV observations reported here and other 
multiwavelength and multimode observations available in the literature, 
the optical spectra of them do not show any 
resemblance to that of blazar class of AGN with relativistic jets. 
Seyferts in general have spiral host galaxies. 
Optical imaging observations of 1H 0323+342 
shows a ring like structure, which hints
of a possible collision with another galaxy (\citealt{2008A&A...490..583A}).
Such interaction with another galaxy could trigger an AGN 
activity (\citealt{2007MNRAS.375.1017A}).
Also, the images obtained from HST Wide Field Planetary Camera 
using the F702W filter corresponding to $\lambda_{eff}$ = 6919 \AA
~is well represented when decomposed with a central
point source plus a S\'{e}rsic component (\citealt{2007ApJ...658L..13Z}).
If the other two sources 
are also conclusively found to be hosted in spiral galaxies, then it points 
to a rethink of the well known paradigm of jets being associated to 
elliptical galaxies.
Further dedicated flux, and optical  polarization  monitoring observations 
coupled
with high resolution optical imaging studies will give clues to the nature
of this new class of $\gamma$-ray loud NLSy1 galaxies. 

\section*{Acknowledgments}
The authors thank the anonymous referee for his/her critical reviewing
and constructive suggestions that helped to improve the presentation.

\def\aj{AJ}%
\def\actaa{Acta Astron.}%
\def\araa{ARA\&A}%
\def\apj{ApJ}%
\def\apjl{ApJ}%
\def\apjs{ApJS}%
\def\ao{Appl.~Opt.}%
\def\apss{Ap\&SS}%
\def\aap{A\&A}%
\def\aapr{A\&A~Rev.}%
\def\aaps{A\&AS}%
\def\azh{AZh}%
\def\baas{BAAS}%
\def\bac{Bull. astr. Inst. Czechosl.}%
\def\caa{Chinese Astron. Astrophys.}%
\def\cjaa{Chinese J. Astron. Astrophys.}%
\def\icarus{Icarus}%
\def\jcap{J. Cosmology Astropart. Phys.}%
\def\jrasc{JRASC}%
\def\mnras{MNRAS}%
\def\memras{MmRAS}%
\def\na{New A}%
\def\nar{New A Rev.}%
\def\pasa{PASA}%
\def\pra{Phys.~Rev.~A}%
\def\prb{Phys.~Rev.~B}%
\def\prc{Phys.~Rev.~C}%
\def\prd{Phys.~Rev.~D}%
\def\pre{Phys.~Rev.~E}%
\def\prl{Phys.~Rev.~Lett.}%
\def\pasp{PASP}%
\def\pasj{PASJ}%
\def\qjras{QJRAS}%
\def\rmxaa{Rev. Mexicana Astron. Astrofis.}%
\def\skytel{S\&T}%
\def\solphys{Sol.~Phys.}%
\def\sovast{Soviet~Ast.}%
\def\ssr{Space~Sci.~Rev.}%
\def\zap{ZAp}%
\def\nat{Nature}%
\def\iaucirc{IAU~Circ.}%
\def\aplett{Astrophys.~Lett.}%
\def\apspr{Astrophys.~Space~Phys.~Res.}%
\def\bain{Bull.~Astron.~Inst.~Netherlands}%
\def\fcp{Fund.~Cosmic~Phys.}%
\def\gca{Geochim.~Cosmochim.~Acta}%
\def\grl{Geophys.~Res.~Lett.}%
\def\jcp{J.~Chem.~Phys.}%
\def\jgr{J.~Geophys.~Res.}%
\def\jqsrt{J.~Quant.~Spec.~Radiat.~Transf.}%
\def\memsai{Mem.~Soc.~Astron.~Italiana}%
\def\nphysa{Nucl.~Phys.~A}%
\def\physrep{Phys.~Rep.}%
\def\physscr{Phys.~Scr}%
\def\planss{Planet.~Space~Sci.}%
\def\procspie{Proc.~SPIE}%
\let\astap=\aap
\let\apjlett=\apjl
\let\apjsupp=\apjs
\let\applopt=\ao
\bibliographystyle{mn}
\bibliography{mybib}

\begin{thebibliography}{57}
\expandafter\ifx\csname natexlab\endcsname\relax\def\natexlab#1{#1}\fi

\bibitem[{{Abdo} {et~al.}(2010){Abdo}, {Ackermann}, {Ajello}, {Allafort},
  {Antolini}, {Atwood}, {Axelsson}, {Baldini}, {Ballet}, {Barbiellini},
  {Bastieri}, {Baughman}, {Bechtol}, {Bellazzini}, {Berenji}, {Blandford},
  {Bloom}, {Bogart}, {Bonamente}, {Borgland}, {Bouvier}, {Bregeon}, {Brez},
  {Brigida}, {Bruel}, {Buehler}, {Burnett}, {Buson}, {Caliandro}, {Cameron},
  {Cannon}, {Caraveo}, {Carrigan}, {Casandjian}, {Cavazzuti}, {Cecchi}, {{\c
  C}elik}, {Celotti}, {Charles}, {Chekhtman}, {Chen}, {Cheung}, {Chiang},
  {Ciprini}, {Claus}, {Cohen-Tanugi}, {Conrad}, {Costamante}, {Cotter},
  {Cutini}, {D'Elia}, {Dermer}, {de Angelis}, {de Palma}, {De Rosa}, {Digel},
  {Silva}, {Drell}, {Dubois}, {Dumora}, {Escande}, {Farnier}, {Favuzzi},
  {Fegan}, {Ferrara}, {Focke}, {Fortin}, {Frailis}, {Fukazawa}, {Funk},
  {Fusco}, {Gargano}, {Gasparrini}, {Gehrels}, {Germani}, {Giebels},
  {Giglietto}, {Giommi}, {Giordano}, {Giroletti}, {Glanzman}, {Godfrey},
  {Grandi}, {Grenier}, {Grondin}, {Grove}, {Guiriec}, {Hadasch}, {Harding},
  {Hayashida}, {Hays}, {Healey}, {Hill}, {Horan}, {Hughes}, {Iafrate}, {Itoh},
  {J{\'o}hannesson}, {Johnson}, {Johnson}, {Johnson}, {Johnson}, {Kamae},
  {Katagiri}, {Kataoka}, {Kawai}, {Kerr}, {Kn{\"o}dlseder}, {Kuss}, {Lande},
  {Latronico}, {Lavalley}, {Lemoine-Goumard}, {Llena Garde}, {Longo},
  {Loparco}, {Lott}, {Lovellette}, {Lubrano}, {Madejski}, {Makeev}, {Malaguti},
  {Massaro}, {Mazziotta}, {McConville}, {McEnery}, {McGlynn}, {Michelson},
  {Mitthumsiri}, {Mizuno}, {Moiseev}, {Monte}, {Monzani}, {Morselli},
  {Moskalenko}, {Murgia}, {Nolan}, {Norris}, {Nuss}, {Ohno}, {Ohsugi},
  {Omodei}, {Orlando}, {Ormes}, {Ozaki}, {Paneque}, {Panetta}, {Parent},
  {Pelassa}, {Pepe}, {Pesce-Rollins}, {Piranomonte}, {Piron}, {Porter},
  {Rain{\`o}}, {Rando}, {Razzano}, {Reimer}, {Reimer}, {Reposeur}, {Ripken},
  {Ritz}, {Rodriguez}, {Romani}, {Roth}, {Ryde}, {Sadrozinski}, {Sanchez},
  {Sander}, {Saz Parkinson}, {Scargle}, {Sgr{\`o}}, {Shaw}, {Siskind}, {Smith},
  {Spandre}, {Spinelli}, {Starck}, {Stawarz}, {Strickman}, {Suson}, {Tajima},
  {Takahashi}, {Takahashi}, {Tanaka}, {Taylor}, {Thayer}, {Thayer}, {Thompson},
  {Tibaldo}, {Torres}, {Tosti}, {Tramacere}, {Ubertini}, {Uchiyama}, {Usher},
  {Vasileiou}, {Vilchez}, {Villata}, {Vitale}, {Waite}, {Wallace}, {Wang},
  {Winer}, {Wood}, {Yang}, {Ylinen}, \& {Ziegler}}]{2010ApJ...715..429A}
{Abdo}, A.~A., {Ackermann}, M., {Ajello}, M., {et~al.}, 2010, \apj, 715, 429

\bibitem[{{Abdo} {et~al.}(2009{\natexlab{a}}){Abdo}, {Ackermann}, {Ajello},
  {Axelsson}, {Baldini}, {Ballet}, {Barbiellini}, {Bastieri}, {Battelino},
  {Baughman}, {Bechtol}, {Bellazzini}, {Bloom}, {Bonamente}, {Borgland},
  {Bregeon}, {Brez}, {Brigida}, {Bruel}, {Caliandro}, {Cameron}, {Caraveo},
  {Casandjian}, {Cavazzuti}, {Cecchi}, {Chekhtman}, {Cheung}, {Chiang},
  {Ciprini}, {Claus}, {Cohen-Tanugi}, {Collmar}, {Conrad}, {Costamante},
  {Dermer}, {de Angelis}, {de Palma}, {Digel}, {Silva}, {Drell}, {Dubois},
  {Dumora}, {Farnier}, {Favuzzi}, {Focke}, {Foschini}, {Frailis}, {Fuhrmann},
  {Fukazawa}, {Funk}, {Fusco}, {Gargano}, {Gehrels}, {Germani}, {Giebels},
  {Giglietto}, {Giordano}, {Giroletti}, {Glanzman}, {Grenier}, {Grondin},
  {Grove}, {Guillemot}, {Guiriec}, {Hanabata}, {Harding}, {Hartman},
  {Hayashida}, {Hays}, {Hughes}, {J{\'o}hannesson}, {Johnson}, {Johnson},
  {Johnson}, {Kamae}, {Katagiri}, {Kataoka}, {Kerr}, {Kn{\"o}dlseder}, {Kuehn},
  {Kuss}, {Lande}, {Latronico}, {Lemoine-Goumard}, {Longo}, {Loparco}, {Lott},
  {Lovellette}, {Lubrano}, {Madejski}, {Makeev}, {Max-Moerbeck}, {Mazziotta},
  {McConville}, {McEnery}, {Meurer}, {Michelson}, {Mitthumsiri}, {Mizuno},
  {Monte}, {Monzani}, {Morselli}, {Moskalenko}, {Murgia}, {Nolan}, {Norris},
  {Nuss}, {Ohsugi}, {Omodei}, {Orlando}, {Ormes}, {Paneque}, {Panetta},
  {Parent}, {Pavlidou}, {Pearson}, {Pepe}, {Pesce-Rollins}, {Piron}, {Porter},
  {Rain{\`o}}, {Rando}, {Razzano}, {Readhead}, {Reimer}, {Reimer}, {Reposeur},
  {Richards}, {Ritz}, {Rodriguez}, {Romani}, {Ryde}, {Sadrozinski}, {Sambruna},
  {Sanchez}, {Sander}, {Parkinson}, {Scargle}, {Schalk}, {Sgr{\`o}}, {Smith},
  {Spandre}, {Spinelli}, {Starck}, {Stevenson}, {Strickman}, {Suson},
  {Tagliaferri}, {Takahashi}, {Tanaka}, {Thayer}, {Thompson}, {Tibaldo},
  {Tibolla}, {Torres}, {Tosti}, {Tramacere}, {Uchiyama}, {Usher}, {Vilchez},
  {Vitale}, {Waite}, {Winer}, {Wood}, {Ylinen}, {Zensus}, {Ziegler}, {Fermi/LAT
  Collaboration}, {Ghisellini}, {Maraschi}, {Tavecchio}, \&
  {Angelakis}}]{2009ApJ...699..976A}
---, 2009{\natexlab{a}}, \apj, 699, 976

\bibitem[{{Abdo} {et~al.}(2009{\natexlab{b}}){Abdo}, {Ackermann}, {Ajello},
  {Axelsson}, {Baldini}, {Ballet}, {Barbiellini}, {Bastieri}, {Baughman},
  {Bechtol}, \& et~al.}]{2009ApJ...707..727A}
---, 2009{\natexlab{b}}, \apj, 707, 727

\bibitem[{{Abdo} {et~al.}(2009{\natexlab{c}}){Abdo}, {Ackermann}, {Ajello},
  {Baldini}, {Ballet}, {Barbiellini}, {Bastieri}, {Bechtol}, {Bellazzini},
  {Berenji}, {Bloom}, {Bonamente}, {Borgland}, {Bregeon}, {Brez}, {Brigida},
  {Bruel}, {Burnett}, {Caliandro}, {Cameron}, {Caraveo}, {Casandjian},
  {Cecchi}, {{\c C}elik}, {Chekhtman}, {Cheung}, {Chiang}, {Ciprini}, {Claus},
  {Cohen-Tanugi}, {Conrad}, {Cutini}, {Dermer}, {de Palma}, {Silva}, {Drell},
  {Dubois}, {Dumora}, {Farnier}, {Favuzzi}, {Fegan}, {Focke}, {Foschini},
  {Frailis}, {Fukazawa}, {Fusco}, {Gargano}, {Gehrels}, {Germani}, {Giebels},
  {Giglietto}, {Giordano}, {Giroletti}, {Glanzman}, {Godfrey}, {Grenier},
  {Grove}, {Guillemot}, {Guiriec}, {Hayashida}, {Hays}, {Horan}, {Hughes},
  {J{\'o}hannesson}, {Johnson}, {Johnson}, {Kadler}, {Kamae}, {Katagiri},
  {Kataoka}, {Kerr}, {Kn{\"o}dlseder}, {Kuss}, {Lande}, {Latronico}, {Longo},
  {Loparco}, {Lott}, {Lovellette}, {Lubrano}, {Makeev}, {Mazziotta},
  {McConville}, {McEnery}, {Meurer}, {Michelson}, {Mitthumsiri}, {Mizuno},
  {Monte}, {Monzani}, {Morselli}, {Moskalenko}, {Murgia}, {Nolan}, {Norris},
  {Nuss}, {Ohsugi}, {Omodei}, {Orlando}, {Ormes}, {Pelassa}, {Pepe}, {Persic},
  {Pesce-Rollins}, {Piron}, {Porter}, {Rain{\`o}}, {Rando}, {Razzano},
  {Rochester}, {Rodriguez}, {Ryde}, {Sadrozinski}, {Sambruna}, {Sander}, {Saz
  Parkinson}, {Scargle}, {Sgr{\`o}}, {Smith}, {Spandre}, {Spinelli},
  {Strickman}, {Suson}, {Tagliaferri}, {Takahashi}, {Takahashi}, {Tanaka},
  {Thayer}, {Thayer}, {Thompson}, {Tibaldo}, {Tibolla}, {Torres}, {Tosti},
  {Tramacere}, {Uchiyama}, {Usher}, {Vasileiou}, {Vilchez}, {Vitale}, {Waite},
  {Wang}, {Winer}, {Wood}, {Ylinen}, {Ziegler}, {Fermi/LAT Collaboration},
  {Ghisellini}, {Maraschi}, \& {Tavecchio}}]{2009ApJ...707L.142A}
---, 2009{\natexlab{c}}, \apjl, 707, L142

\bibitem[{{Alonso} {et~al.}(2007){Alonso}, {Lambas}, {Tissera}, \&
  {Coldwell}}]{2007MNRAS.375.1017A}
{Alonso}, M.~S., {Lambas}, D.~G., {Tissera}, P., \& {Coldwell}, G., 2007,
  \mnras, 375, 1017

\bibitem[{{Ant{\'o}n} {et~al.}(2008){Ant{\'o}n}, {Browne}, \&
  {March{\~a}}}]{2008A&A...490..583A}
{Ant{\'o}n}, S., {Browne}, I.~W.~A., \& {March{\~a}}, M.~J., 2008, \aap, 490,
  583

\bibitem[{{Arshakian} {et~al.}(2010){Arshakian}, {Le{\'o}n-Tavares}, {Lobanov},
  {Chavushyan}, {Shapovalova}, {Burenkov}, \& {Zensus}}]{2010MNRAS.401.1231A}
{Arshakian}, T.~G., {Le{\'o}n-Tavares}, J., {Lobanov}, A.~P., {Chavushyan},
  V.~H., {Shapovalova}, A.~I., {Burenkov}, A.~N., \& {Zensus}, J.~A., 2010,
  \mnras, 401, 1231

\bibitem[{{Begelman} {et~al.}(1984){Begelman}, {Blandford}, \&
  {Rees}}]{1984RvMP...56..255B}
{Begelman}, M.~C., {Blandford}, R.~D., \& {Rees}, M.~J., 1984, Reviews of
  Modern Physics, 56, 255

\bibitem[{{Carini}(1990)}]{1990BAAS...22R1337C}
{Carini}, M.~T., 1990, in Bulletin of the American Astronomical Society,
  Vol.~22, Bulletin of the American Astronomical Society, p. 1337

\bibitem[{{Carini} {et~al.}(1990){Carini}, {Miller}, \&
  {Goodrich}}]{1990AJ....100..347C}
{Carini}, M.~T., {Miller}, H.~R., \& {Goodrich}, B.~D., 1990, \aj, 100, 347

\bibitem[{{Cellone} {et~al.}(2000){Cellone}, {Romero}, \&
  {Combi}}]{2000AJ....119.1534C}
{Cellone}, S.~A., {Romero}, G.~E., \& {Combi}, J.~A., 2000, \aj, 119, 1534

\bibitem[{{D'Ammando} {et~al.}(2012){D'Ammando}, {Orienti}, {Finke}, {Raiteri},
  {Angelakis}, {Fuhrmann}, {Giroletti}, {Hovatta}, {Max-Moerbeck}, {Perkins},
  {Readhead}, {Richards}, {Stawarz}, \& {Donato}}]{2012arXiv1207.3092D}
{D'Ammando}, F., {Orienti}, M., {Finke}, J., {et~al.}, 2012, ArXiv e-prints

\bibitem[{{de Diego}(2010)}]{2010AJ....139.1269D}
{de Diego}, J.~A., 2010, \aj, 139, 1269

\bibitem[{{Doi} {et~al.}(2006){Doi}, {Nagai}, {Asada}, {Kameno}, {Wajima}, \&
  {Inoue}}]{2006PASJ...58..829D}
{Doi}, A., {Nagai}, H., {Asada}, K., {Kameno}, S., {Wajima}, K., \& {Inoue},
  M., 2006, \pasj, 58, 829

\bibitem[{{Eggen} {et~al.}(2011){Eggen}, {Miller}, \&
  {Maune}}]{2011nlsg.confE..49E}
{Eggen}, J., {Miller}, H.~R., \& {Maune}, J., 2011, in Narrow-Line Seyfert 1
  Galaxies and their Place in the Universe

\bibitem[{{Foschini}(2011)}]{2011nlsg.confE..24F}
{Foschini}, L., 2011, in Narrow-Line Seyfert 1 Galaxies and their Place in the
  Universe

\bibitem[{{Foschini} {et~al.}(2010){Foschini}, {Fermi/Lat Collaboration},
  {Ghisellini}, {Maraschi}, {Tavecchio}, \& {Angelakis}}]{2010ASPC..427..243F}
{Foschini}, L., {Fermi/Lat Collaboration}, {Ghisellini}, G., {Maraschi}, L.,
  {Tavecchio}, F., \& {Angelakis}, E., 2010, in Astronomical Society of the
  Pacific Conference Series, Vol. 427, Accretion and Ejection in AGN: a Global
  View, {Maraschi}, L., {Ghisellini}, G., {Della Ceca}, R., \& {Tavecchio}, F.,
  eds., pp. 243--248

\bibitem[{{Foschini} {et~al.}(2011){Foschini}, {Ghisellini}, {Kovalev},
  {Lister}, {D'Ammando}, {Thompson}, {Tramacere}, {Angelakis}, {Donato},
  {Falcone}, {Fuhrmann}, {Hauser}, {Kovalev}, {Mannheim}, {Maraschi},
  {Max-Moerbeck}, {Nestoras}, {Pavlidou}, {Pearson}, {Pushkarev}, {Readhead},
  {Richards}, {Stevenson}, {Tagliaferri}, {Tibolla}, {Tavecchio}, \&
  {Wagner}}]{2011MNRAS.413.1671F}
{Foschini}, L., {Ghisellini}, G., {Kovalev}, Y.~Y., {et~al.}, 2011, \mnras,
  413, 1671

\bibitem[{{Giroletti} {et~al.}(2011){Giroletti}, {Paragi}, {Bignall}, {Doi},
  {Foschini}, {Gab{\'a}nyi}, {Reynolds}, {Blanchard}, {Campbell}, {Colomer},
  {Hong}, {Kadler}, {Kino}, {van Langevelde}, {Nagai}, {Phillips}, {Sekido},
  {Szomoru}, \& {Tzioumis}}]{2011A&A...528L..11G}
{Giroletti}, M., {Paragi}, Z., {Bignall}, H., {et~al.}, 2011, \aap, 528, L11

\bibitem[{{Gopal-Krishna} {et~al.}(2011){Gopal-Krishna}, {Goyal}, {Joshi},
  {Karthick}, {Sagar}, {Wiita}, {Anupama}, \& {Sahu}}]{2011MNRAS.416..101G}
{Gopal-Krishna}, {Goyal}, A., {Joshi}, S., {Karthick}, C., {Sagar}, R.,
  {Wiita}, P.~J., {Anupama}, G.~C., \& {Sahu}, D.~K., 2011, \mnras, 416, 101

\bibitem[{{Goyal} {et~al.}(2012){Goyal}, {Gopal-Krishna}, {Wiita}, {Anupama},
  {Sahu}, {Sagar}, \& {Joshi}}]{2012A&A...544A..37G}
{Goyal}, A., {Gopal-Krishna}, {Wiita}, P.~J., {Anupama}, G.~C., {Sahu}, D.~K.,
  {Sagar}, R., \& {Joshi}, S., 2012, \aap, 544, A37

\bibitem[{{Hagen-Thorn} {et~al.}(2008){Hagen-Thorn}, {Larionov}, {Jorstad},
  {Arkharov}, {Hagen-Thorn}, {Efimova}, {Larionova}, \&
  {Marscher}}]{2008ApJ...672...40H}
{Hagen-Thorn}, V.~A., {Larionov}, V.~M., {Jorstad}, S.~G., {Arkharov}, A.~A.,
  {Hagen-Thorn}, E.~I., {Efimova}, N.~V., {Larionova}, L.~V., \& {Marscher},
  A.~P., 2008, \apj, 672, 40

\bibitem[{{Hughes} {et~al.}(1992){Hughes}, {Aller}, \&
  {Aller}}]{1992ApJ...396..469H}
{Hughes}, P.~A., {Aller}, H.~D., \& {Aller}, M.~F., 1992, \apj, 396, 469

\bibitem[{{Ikejiri} {et~al.}(2011){Ikejiri}, {Uemura}, {Sasada}, {Ito},
  {Yamanaka}, {Sakimoto}, {Arai}, {Fukazawa}, {Ohsugi}, {Kawabata}, {Yoshida},
  {Sato}, \& {Kino}}]{2011PASJ...63..639I}
{Ikejiri}, Y., {Uemura}, M., {Sasada}, M., {et~al.}, 2011, \pasj, 63, 639

\bibitem[{{Impiombato} {et~al.}(2011){Impiombato}, {Covino}, {Treves},
  {Foschini}, {Pian}, {Tosti}, {Fugazza}, {Nicastro}, \&
  {Ciprini}}]{2011ApJS..192...12I}
{Impiombato}, D., {Covino}, S., {Treves}, A., {et~al.}, 2011, \apjs, 192, 12

\bibitem[{{Jang} \& {Miller}(1997)}]{1997AJ....114..565J}
{Jang}, M. \& {Miller}, H.~R., 1997, \aj, 114, 565

\bibitem[{{Klimek} {et~al.}(2004){Klimek}, {Gaskell}, \&
  {Hedrick}}]{2004ApJ...609...69K}
{Klimek}, E.~S., {Gaskell}, C.~M., \& {Hedrick}, C.~H., 2004, \apj, 609, 69

\bibitem[{{Komossa} {et~al.}(2006){Komossa}, {Voges}, {Xu}, {Mathur}, {Adorf},
  {Lemson}, {Duschl}, \& {Grupe}}]{2006AJ....132..531K}
{Komossa}, S., {Voges}, W., {Xu}, D., {Mathur}, S., {Adorf}, H.-M., {Lemson},
  G., {Duschl}, W.~J., \& {Grupe}, D., 2006, \aj, 132, 531

\bibitem[{{Laor}(2000)}]{2000ApJ...543L.111L}
{Laor}, A., 2000, \apjl, 543, L111

\bibitem[{{Le{\'o}n-Tavares} {et~al.}(2010){Le{\'o}n-Tavares}, {Lobanov},
  {Chavushyan}, {Arshakian}, {Doroshenko}, {Sergeev}, {Efimov}, \&
  {Nazarov}}]{2010ApJ...715..355L}
{Le{\'o}n-Tavares}, J., {Lobanov}, A.~P., {Chavushyan}, V.~H., {Arshakian},
  T.~G., {Doroshenko}, V.~T., {Sergeev}, S.~G., {Efimov}, Y.~S., \& {Nazarov},
  S.~V., 2010, \apj, 715, 355

\bibitem[{{Liu} {et~al.}(2010){Liu}, {Wang}, {Mao}, \&
  {Wei}}]{2010ApJ...715L.113L}
{Liu}, H., {Wang}, J., {Mao}, Y., \& {Wei}, J., 2010, \apjl, 715, L113

\bibitem[{{Marscher} \& {Gear}(1985)}]{1985ApJ...298..114M}
{Marscher}, A.~P. \& {Gear}, W.~K., 1985, \apj, 298, 114

\bibitem[{{Marscher} {et~al.}(1992){Marscher}, {Gear}, \&
  {Travis}}]{1992vob..conf...85M}
{Marscher}, A.~P., {Gear}, W.~K., \& {Travis}, J.~P., 1992, in Variability of
  Blazars, {Valtaoja}, E. \& {Valtonen}, M., eds., p.~85

\bibitem[{{Mattox} {et~al.}(1996){Mattox}, {Bertsch}, {Chiang}, {Dingus},
  {Digel}, {Esposito}, {Fierro}, {Hartman}, {Hunter}, {Kanbach}, {Kniffen},
  {Lin}, {Macomb}, {Mayer-Hasselwander}, {Michelson}, {von Montigny},
  {Mukherjee}, {Nolan}, {Ramanamurthy}, {Schneid}, {Sreekumar}, {Thompson}, \&
  {Willis}}]{1996ApJ...461..396M}
{Mattox}, J.~R., {Bertsch}, D.~L., {Chiang}, J., {et~al.}, 1996, \apj, 461, 396

\bibitem[{{Maune} {et~al.}(2011){Maune}, {Miller}, \&
  {Eggen}}]{2011nlsg.confE..59M}
{Maune}, J.~D., {Miller}, H.~R., \& {Eggen}, J.~R., 2011, in Narrow-Line
  Seyfert 1 Galaxies and their Place in the Universe

\bibitem[{{Miller} {et~al.}(1989){Miller}, {Carini}, \&
  {Goodrich}}]{1989Natur.337..627M}
{Miller}, H.~R., {Carini}, M.~T., \& {Goodrich}, B.~D., 1989, \nat, 337, 627

\bibitem[{{Miller} {et~al.}(2000){Miller}, {Ferrara}, {McFarland}, {Wilson},
  {Daya}, \& {Fried}}]{2000NewAR..44..539M}
{Miller}, H.~R., {Ferrara}, E.~C., {McFarland}, J.~P., {Wilson}, J.~W., {Daya},
  A.~B., \& {Fried}, R.~E., 2000, \nar, 44, 539

\bibitem[{{Monet} {et~al.}(2003){Monet}, {Levine}, {Canzian}, {Ables}, {Bird},
  {Dahn}, {Guetter}, {Harris}, {Henden}, {Leggett}, {Levison}, {Luginbuhl},
  {Martini}, {Monet}, {Munn}, {Pier}, {Rhodes}, {Riepe}, {Sell}, {Stone},
  {Vrba}, {Walker}, {Westerhout}, {Brucato}, {Reid}, {Schoening}, {Hartley},
  {Read}, \& {Tritton}}]{2003AJ....125..984M}
{Monet}, D.~G., {Levine}, S.~E., {Canzian}, B., {et~al.}, 2003, \aj, 125, 984

\bibitem[{{Noble}(1995)}]{1995PhDT.........2N}
{Noble}, J.~C., 1995, PhD thesis, GEORGIA STATE UNIVERSITY.

\bibitem[{{Orienti} {et~al.}(2012){Orienti}, {D'Ammando}, {Giroletti}, \& {for
  the Fermi-LAT Collaboration}}]{2012arXiv1205.0402O}
{Orienti}, M., {D'Ammando}, F., {Giroletti}, M., \& {for the Fermi-LAT
  Collaboration}, 2012, ArXiv e-prints

\bibitem[{{Osterbrock} \& {Pogge}(1985)}]{1985ApJ...297..166O}
{Osterbrock}, D.~E. \& {Pogge}, R.~W., 1985, \apj, 297, 166

\bibitem[{{Pogge}(2011)}]{2011nlsg.confE...2P}
{Pogge}, R.~W., 2011, in Narrow-Line Seyfert 1 Galaxies and their Place in the
  Universe

\bibitem[{{Pushkarev} {et~al.}(2009){Pushkarev}, {Kovalev}, {Lister}, \&
  {Savolainen}}]{2009A&A...507L..33P}
{Pushkarev}, A.~B., {Kovalev}, Y.~Y., {Lister}, M.~L., \& {Savolainen}, T.,
  2009, \aap, 507, L33

\bibitem[{{Rani} {et~al.}(2010){Rani}, {Gupta}, {Joshi}, {Ganesh}, \&
  {Wiita}}]{2010ApJ...719L.153R}
{Rani}, B., {Gupta}, A.~C., {Joshi}, U.~C., {Ganesh}, S., \& {Wiita}, P.~J.,
  2010, \apjl, 719, L153

\bibitem[{{Rodr{\'{\i}}guez-Ardila} {et~al.}(2000){Rodr{\'{\i}}guez-Ardila},
  {Binette}, {Pastoriza}, \& {Donzelli}}]{2000ApJ...538..581R}
{Rodr{\'{\i}}guez-Ardila}, A., {Binette}, L., {Pastoriza}, M.~G., \&
  {Donzelli}, C.~J., 2000, \apj, 538, 581

\bibitem[{{Romero} {et~al.}(1999){Romero}, {Cellone}, \&
  {Combi}}]{1999A&AS..135..477R}
{Romero}, G.~E., {Cellone}, S.~A., \& {Combi}, J.~A., 1999, \aaps, 135, 477

\bibitem[{{Sagar} {et~al.}(2012){Sagar}, B., {Omar}, \& Y.}]{2012aj}
{Sagar}, R., B., K., {Omar}, A., \& Y., J., 2012, ASI Conference Series, 4, 113

\bibitem[{{Sagar} {et~al.}(2010){Sagar}, {Kumar}, {Omar}, \&
  {Pandey}}]{2010ASInC...1..203S}
{Sagar}, R., {Kumar}, B., {Omar}, A., \& {Pandey}, A.~K., 2010, in Astronomical
  Society of India Conference Series, Vol.~1, Astronomical Society of India
  Conference Series, pp. 203--210

\bibitem[{{Sagar} {et~al.}(2004){Sagar}, {Stalin}, {Gopal-Krishna}, \&
  {Wiita}}]{2004MNRAS.348..176S}
{Sagar}, R., {Stalin}, C.~S., {Gopal-Krishna}, \& {Wiita}, P.~J., 2004, \mnras,
  348, 176

\bibitem[{{Stalin} {et~al.}(2004){Stalin}, {Gopal Krishna}, {Sagar}, \&
  {Wiita}}]{2004JApA...25....1S}
{Stalin}, C.~S., {Gopal Krishna}, {Sagar}, R., \& {Wiita}, P.~J., 2004, Journal
  of Astrophysics and Astronomy, 25, 1

\bibitem[{{Tosti} {et~al.}(1998){Tosti}, {Fiorucci}, {Luciani}, {Efimov},
  {Shakhovskoy}, {Valtaoja}, {Teraesranta}, {Sillanpaeae}, {Takalo}, {Villata},
  {Raiteri}, {de Francesco}, \& {Sobrito}}]{1998A&A...339...41T}
{Tosti}, G., {Fiorucci}, M., {Luciani}, M., {et~al.}, 1998, \aap, 339, 41

\bibitem[{{Urry} \& {Padovani}(1995)}]{1995PASP..107..803U}
{Urry}, C.~M. \& {Padovani}, P., 1995, \pasp, 107, 803

\bibitem[{{V{\'e}ron-Cetty} \& {V{\'e}ron}(2010)}]{2010A&A...518A..10V}
{V{\'e}ron-Cetty}, M.-P. \& {V{\'e}ron}, P., 2010, \aap, 518, A10

\bibitem[{{Wagner} \& {Witzel}(1995)}]{1995ARA&A..33..163W}
{Wagner}, S.~J. \& {Witzel}, A., 1995, \araa, 33, 163

\bibitem[{{Webb} \& {Malkan}(2000)}]{2000ApJ...540..652W}
{Webb}, W. \& {Malkan}, M., 2000, \apj, 540, 652

\bibitem[{{Zhou} {et~al.}(2007){Zhou}, {Wang}, {Yuan}, {Shan}, {Komossa}, {Lu},
  {Liu}, {Xu}, {Bai}, \& {Jiang}}]{2007ApJ...658L..13Z}
{Zhou}, H., {Wang}, T., {Yuan}, W., {et~al.}, 2007, \apjl, 658, L13

\bibitem[{{Zhou} {et~al.}(2003){Zhou}, {Wang}, {Dong}, {Zhou}, \&
  {Li}}]{2003ApJ...584..147Z}
{Zhou}, H.-Y., {Wang}, T.-G., {Dong}, X.-B., {Zhou}, Y.-Y., \& {Li}, C., 2003,
  \apj, 584, 147

\end{thebibliography}
\end{document}